\documentclass[proceedings]{ccn}  
\ccndoi{10.32470/0416gfsq}

\usepackage{caption}
\usepackage{booktabs}
\usepackage{multirow}
\usepackage{tabularx}
\usepackage{ragged2e,array}
\usepackage{amsmath}
\usepackage[utf8]{inputenc}
\usepackage[T1]{fontenc}
\usepackage{placeins}
\usepackage{float}
\usepackage{siunitx}
\usepackage{makecell}
\usepackage{graphicx}
\usepackage{subcaption}
\usepackage{csvsimple}

\usepackage{xcolor}

\title{Eccentricity-Constrained CNN Training Reveals Adaptive Information Coding Around the Visual Field}

\author{Dylan M. Diaz \\
  Departments of Computer Science and Psychology, California State University, San Bernardino \\
  \email{dylan.diaz4811@coyote.csusb.edu}
  \AND Margaret M. Henderson \\
  Department of Psychology and Neuroscience Institute, Carnegie Mellon University \\
  \email{mmhender@andrew.cmu.edu}}

\begin{document}
\raggedbottom
\maketitle

\begin{abstract}
Within topographic eccentricity maps in the primate visual system, center-preferring populations have higher spatial resolution and overlap face- and word-selective regions while periphery-preferring populations have lower spatial resolution and overlap scene-selective regions. Prior behavioral and neuroimaging evidence suggests that this ``eccentricity bias’’ may reflect the relevance of visual field portions for different tasks: the central visual field may be more informative for fine-grained tasks like face recognition and reading, while the periphery may be more informative for large-scale scene understanding tasks. To examine whether such eccentricity-dependent coding can emerge from natural experience, we leveraged egocentric video and eye-tracking data from the Visual Experience Dataset (VEDB). We trained ResNet-18 models using contrastive learning (SimCLR) on video frames modified to isolate information available at different eccentricities (gaze-contingent fovea-only crops, periphery-only crops, and periphery-only crops with a NeuroFovea transform applied). We then evaluated downstream task performance and model alignment with human fMRI data (Natural Scenes Dataset; encoding model framework). When examining in-domain classification of VEDB frame categories, we observed systematic variability in the performance of fovea-only and periphery-only models across categories, suggesting differential informativeness of visual field eccentricities across tasks. 
On downstream classification without fine-tuning, VEDB-pretrained models
generalized more strongly to scene recognition (Places365) than to face recognition (VGGFace2), with fovea-only models showing an advantage on both tasks.
Across visual cortex, VEDB-pretrained models achieved similar neural predictivity to models trained on mid-sized non-egocentric datasets (ImageNet-100), suggesting experience-sampled egocentric data, despite its low diversity and constrained semantic content, supports emergence of cortically-aligned representations.
In scene-selective sub-regions (PPA, RSC), periphery-only models held a small but consistent advantage in explained variance over fovea-only models, suggesting scene-selective cortex may be adapted to peripheral-field statistics.
Together, these results suggest that naturalistic egocentric experience provides an organizing constraint on perception, leading to adaptive, task-aligned information processing.
\end{abstract}


\begin{figure*}[t]
  \centering
  \makebox[\textwidth][c]{%
    \includegraphics[width=1.00\textwidth]{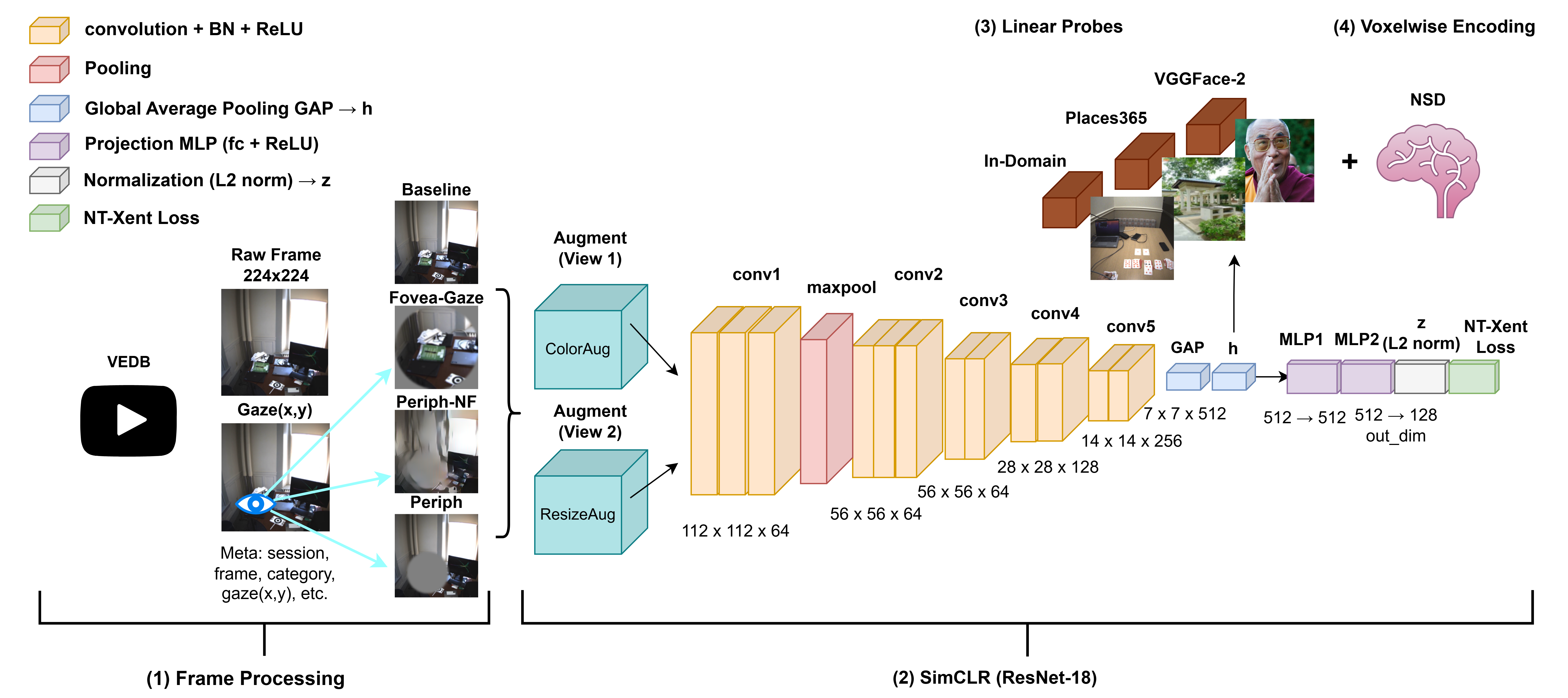}%
  }
  \caption{Method overview. VEDB frames and gaze metadata are processed to create eccentricity-constrained inputs (Baseline, Fovea-Gaze, Periph-NF, Periph). Fovea-Gaze uses 112×112 crops centered on per-frame gaze position. Condition-specific SimCLR pretraining uses a Siamese architecture with tied (shared) ResNet-18 weights across the two augmented views; see \textit{Methods}.
  Downstream evaluation uses frozen-backbone linear probes (in-domain, VGGFace2, Places365) and voxel-wise encoding on NSD.}
  \label{fig:method_overview}
\end{figure*}

\section{Introduction}
Visual field eccentricity is a key organizing dimension of the primate visual system, with topographic maps of preferred eccentricity found across early and higher visual cortex. These spatial eccentricity maps also align with topographic representations of mid-level features (curvature, spatial frequency) and semantic categories (faces, words, scenes), such that regions coding the center of visual space tend to have higher resolution and be selective for curved contours, as well as overlapping with face-selective and word-selective regions, while regions preferring the periphery tend to overlap with scene-selective visual regions \citep{Levy2001, Hasson2002, Ponce2017, Srihasam2014, Yue2014, Yue2020, Arcaro2017, Henderson2023a}. 

According to one hypothesis, this organization may reflect the differential task-relevance of the central and peripheral visual field portions for different behavioral tasks. The central visual field, with its small spatial receptive fields and high acuity, may be useful for fine-grained tasks like face recognition and reading, while more global tasks like scene recognition may benefit from the larger receptive fields of neurons in peripheral-coding regions \citep{Hasson2002, larson2009contributions, Wang2017, Henderson2023a}. Relatedly, the eccentricity biases found in higher visual cortex in the adult visual system may emerge in part due to the portion of retinotopic space in which face, word, and scene content typically occur (e.g., faces are often foveated, buildings often land in the peripheral field). 

This account suggests that in natural egocentric inputs, image statistics may differ systematically between the central and peripheral portions of the visual field. These differences may include mid-level properties like curvature and spatial frequency, and/or semantic content (e.g., faces or scene components). If the visual system is adapted to these statistics, this
eccentricity-specific coding could have functional benefits for downstream tasks like face and scene recognition. 

Critically, testing whether such eccentricity-dependent coding emerges from natural experience requires large-scale naturalistic egocentric training data that incorporate eye movement behavior.
To achieve this here, we leverage
the Visual Experience Dataset (VEDB; \cite{greene2024visual}) to train eccentricity-constrained convolutional neural network (CNN) models. We extract VEDB frames and modify them to isolate either a fixed gaze-centered crop around the participant’s per-frame gaze location (Fovea-Gaze) or the complementary peripheral region (Periph); we additionally use the NeuroFovea transform (\cite{Deza2019}) to simulate information loss in peripheral vision (Periph-NF). We then pretrain CNNs using self-supervised contrastive learning (SimCLR; \cite{Chen2020}) on each dataset version, and test performance on downstream tasks probing various aspects of scene (Places365; \cite{zhou2017places}) and face (VGGFace2; \cite{Cao2017vgg}) perception. Finally, we evaluate neural predictivity of each model using a large fMRI dataset (Natural Scenes Dataset; NSD; \cite{Allen2022}). Our key contributions are as follows:

\begin{itemize}
\item We show that pretraining with Fovea-Gaze vs. Periph or Periph-NF inputs leads to systematic variability in downstream task performance across categories, providing preliminary evidence that the central and peripheral field are differentially relevant for classifying object-centric versus large-scale scene content. 
\item We demonstrate VEDB-pretrained models generalize more effectively to scene classification (Places365) than face classification (VGGFace2), and that Fovea-Gaze outperforms peripheral models on both tasks. 
{\item We show that self-supervised VEDB pretraining leads to comparable fMRI encoding model prediction accuracy relative to mid-sized non-egocentric datasets (STL-10, ImageNet-100), approaching accuracy of models trained on a large dataset (ImageNet-1K).}
\item We find small but statistically reliable differences in unique variance, with the Periph model favored in scene-selective cortex (PPA, RSC) and the Fovea-Gaze model favored in primary visual cortex.
\end{itemize}

\subsection{Related Work}

Behavioral evidence suggests that object and face recognition are best in the central visual field \citep{henderson1999role, nelson1980functional, pelli2008uncrowded, rosenholtz2016capabilities}, while scene recognition may rely more on peripheral features \citep{mccotter2005visual, larson2009contributions}. The role of peripheral vision in scene recognition may reflect large receptive fields and gist-like structural information, with some models proposing peripheral vision represents information in a texture-like code with large pooling fields \citep{rosenholtz2016capabilities, Freeman2011, rosenholtz2012summary, oliva2006building}. In the higher visual system, information from the central and peripheral field is partially anatomically segregated, with central-preferring regions overlapping face-selective and word-selective areas, and peripheral-preferring regions overlapping with scene-selective areas \citep{grillspector2014functional, Hasson2002, Levy2001}

Computational work has also examined the functional task-relevance of central and peripheral vision. CNN modeling has shown that peripheral features are more useful than central features for scene classification \citep{Wang2017}, and that modeling separate central and peripheral streams benefits downstream task performance 
\citep{Zhang2019}. Other work has
demonstrated simulating cortical magnification and texture-like peripheral encoding can improve model robustness and performance \citep{deza2020emergent, Wang2021}, and that modeling peripheral processing can boost classification performance in a transformer model \citep{min2022peripheral}. Recent work has further connected these ideas to neural encoding and cortical organization, showing that biologically motivated spatial sampling of DNN feature maps improves encoding models of human neural responses
\citep{Muller2024}, and that retinotopically organized inputs can scaffold consistent category-selective topography in models of high-level visual cortex \citep{Blauch2025}.

Much of this previous computational work has used non-egocentric computer vision datasets, whose curation and content coverage may differ strongly from the statistics of naturalistic first-person visual experience \citep{grauman2022ego4dworld3000hours, greene2024visual}. Moreover, this past work does not account for eye gaze position within scenes, 
meaning the retinotopic location of features may not be modeled accurately. Here we fill these gaps by training eccentricity-constrained CNN models on gaze-aligned frames of egocentric video data. Importantly, it is not trivial \textit{a priori }that self-supervised models trained on such data, which is temporally correlated and less diverse than curated image datasets, would learn useful or performant visual representations, relative to curated image datasets. We also provide the novel contribution of directly comparing our models with human fMRI data recorded across early and category-selective cortical regions, enabling a test of the link between egocentric image statistics and cortical representations.

\section{Methods}
\subsection{VEDB Processing and Metamer Generation}
We analyzed egocentric video from the Visual Experience Database (VEDB; \citealt{greene2024visual}), retaining sessions with synchronized gaze. Frames were sampled uniformly every 2 seconds within task-relevant session windows, decoded to RGB, and center-cropped to $224{\times}224$. All splits were performed at the session level to prevent leakage from temporally adjacent frames; SimCLR pretraining used an 80/10/10 train/val/test session split, and in-domain classification was restricted to a core subset of 17 tasks. To approximate peripheral texture-tiling representations, we generated gaze-contingent peripheral metamers using NeuroFovea (NF; \citealt{Deza2019}), a feed-forward foveated style-transfer model \citep{Freeman2011,Rosenholtz2019TTMChallenges}, applied per frame with pooling centered at the per-frame fixation. See Supp. Figure~\ref{fig:supp_frame_examples} for frame examples and \textit{Supp. Methods} for full details.

\begin{figure*}[ht]
  \centering
  \includegraphics[width=\textwidth]{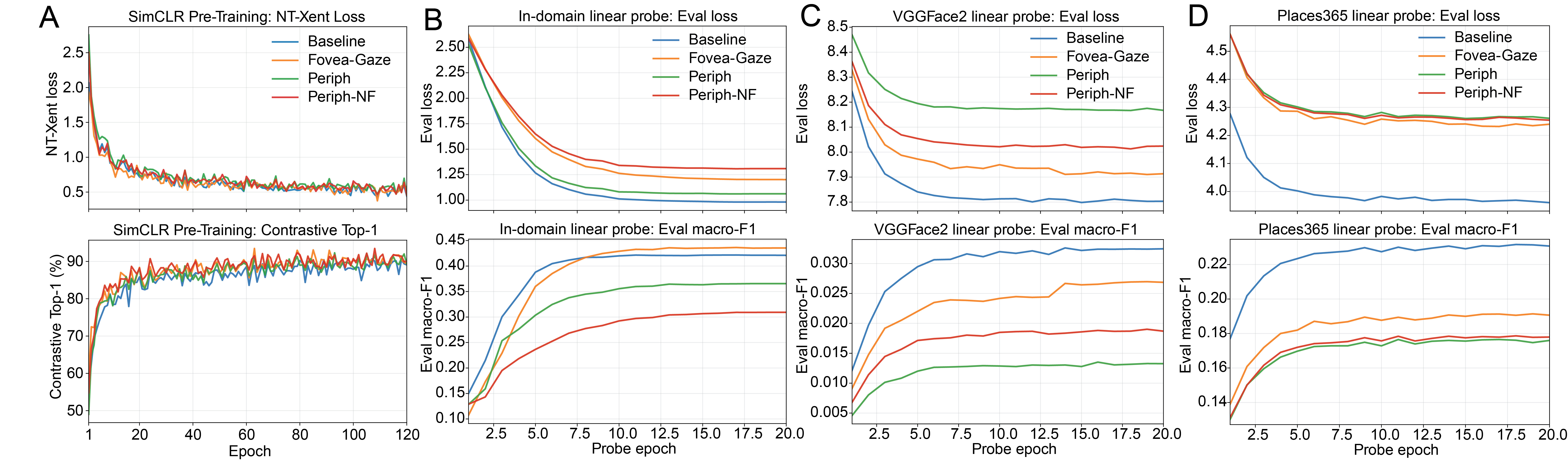}
  \caption{Performance of VEDB models during pretraining and linear probing. \textbf{(A)} SimCLR pre-training curves across conditions (120 epochs). \textbf{(B)} In-domain linear probe learning curves across conditions (20 probe epochs). \textbf{(C)} VGGFace2 transfer linear probe learning curves across conditions (20 probe epochs). \textbf{(D)} Places365 transfer linear probe learning curves across conditions (20 probe epochs).
  }
  \label{fig:curves-combined-all}
\end{figure*}

\subsection{Experimental Setup and Augmentations}

The Baseline, Fovea-Gaze, Periph, and Periph-NF conditions followed an identical two-stage pipeline: (1) SimCLR pre-training on condition-specific augmented inputs, followed by (2) supervised linear-probe evaluation on in-domain classification and transfer tasks (VGGFace2 and Places365 classification).

\subsubsection{1) Baseline.}
Models were trained on the original, unedited 224×224 frames.

\subsubsection{2) Fovea-Gaze.}
We generated Fovea-Gaze views by extracting a fixed $112\times112$ crop from each $224\times224$ frame, centered on the participant's per-frame gaze location (see \textit{Supp. Methods}). Crops were clamped to image bounds and upsampled to $224\times224$ (bilinear interpolation) to maintain a constant input size. To suppress non-crop information, we applied a feathered circular aperture; surrounding region masked with uniform gray (gray=128; Gaussian blur kernel=15); frames without high-confidence gaze used a centered fallback crop. This condition should be interpreted as gaze-centered central-only input, not a simulation of biological fovea.

\subsubsection{3) Periph-NF.}
Periph-NF images were constructed by applying the NeuroFovea transform to each frame prior to occluding the foveal region with a gray circular scotoma (pixel value = 128, Gaussian-blurred mask edges, kernel = 15) centered at the per-frame gaze-defined location. This differs from the Periph condition only in the application of NeuroFovea prior to masking, which introduces eccentricity-dependent texture pooling approximating the reduced acuity of peripheral vision.

\subsubsection{4) Periph.}
Control condition of Periph-NF, constructed identically but without the NeuroFovea transform.



\subsection{SimCLR Pre-Training}
SimCLR pre-training was run once for each condition using a ResNet-18 encoder for 120 epochs (batch size = 64). The augmentation pipeline comprised random resized cropping, horizontal flipping, color jitter, grayscale conversion, and Gaussian blur, and was identical across all conditions, applied after condition-specific frame construction. Full training hyperparameters and projection head architecture are provided in Supp. Table~\ref{tab:supp_hparams_simclr}.

\subsection{Non-VEDB Baseline Models} As comparison baselines, we included ResNet-18 models pretrained with SimCLR on three non-egocentric datasets: STL-10 (\cite{coates2011}), using a pretrained checkpoint (\url{https://github.com/Spijkervet/SimCLR}; 40 epochs); ImageNet-100 (\cite{imagenet100tian2019}; a subset of ImageNet including 100 randomly sampled classes), trained using the LightlySSL codebase, following a standard SimCLR augmentation pipeline for 120 epochs; and ImageNet-1K (\cite{imagenet15russakovsky2015}; images resized to 256 px to reduce computational demand), trained using LightlySSL for 100 epochs.

\subsection{Downstream Classifiers}
For in-domain classification and transfer to VGGFace-2 and Places-365, we trained linear probes on a frozen ResNet-18 backbone initialized from each SimCLR checkpoint. For in-domain evaluation, frames were preprocessed identically to pretraining; no such modifications were applied for transfer. All probes were trained for 20 epochs (AdamW); checkpoints were selected by validation Macro-F1 (in-domain) or Top-1 accuracy (transfer). Hyperparameters and dataset splitting details are summarized in Supp. Table~\ref{tab:supp_hparams_linear_probe_tasks} and \textit{Supp. Methods}. 

\subsection{fMRI Encoding Models}
To compare the ability of eccentricity-constrained pretrained models to predict brain activity across the human visual system, we used a large 7T fMRI dataset consisting of whole-brain responses to complex natural scene images in 8 human participants (NSD; \cite{Allen2022}). We constructed voxelwise encoding models \citep{Naselaris2015} which map from viewed stimuli to single-voxel image-wise activation. Voxel beta values were extracted using GLMSingle \citep{Prince2022} and averaged over repetitions of each image. 

To build encoding models for each participant, we first extracted the model activations from a pretrained ResNet-18 model in response to each stimulus image viewed by the fMRI participant. This was done for each of the Baseline, Fovea-Gaze, Periph, and Periph-NF models separately (following the SimCLR pretraining step).Features were extracted from the original, intact NSD images for all model conditions; condition-specific transforms were applied only during SimCLR pretraining and were not reapplied at the encoding stage. As control comparisons, we also included the non-VEDB models described above (see \textit{Methods} "Non-VEDB Baseline Models").
From each model, we used the output of each of the layers: Conv1, Layer1.1, Layer2.1, Layer3.1, Layer4.1, Avgpool. To reduce dimensionality of features, we used average pooling to spatially downsample convolutional layers, flattened the features, then applied principal components analysis (PCA) and retained the top 200 components. We concatenated features across all layers, z-scored the features across images, and used regularized (L2) linear regression to learn a voxel-specific linear mapping from the model features to the voxel's response, including an intercept term. Encoding model fit was evaluated based on the model's ability to predict responses to held-out images, quantified using $R^2$ on the held-out test set. For each participant, we used the 1000 images in the NSD that were shared across all participants (see \cite{Allen2022} for details) as our test set and fit the model on the remaining 9000 images viewed by the participant.

Functional regions of interest (ROIs) were defined using data from independent localizers. Retinotopic early visual ROIs (V1, V2, V3, hV4) were defined using a receptive field mapping task, and category-selective ROIs (scene-selective: OPA, PPA, RSC; face-selective: FFA, OFA; word-selective: VWFA, OWFA) were defined using a category localizer task (see \cite{Allen2022}). We performed additional thresholding of the category-selective ROI definitions provided by \cite{Allen2022}, retaining only voxels with a localizer test statistic of \textit{t} $>$ 2 for the region's preferred category. Voxels were further thresholded by their estimated noise ceiling (threshold 0.10).

To perform variance partitioning between a pair of models \textit{A} and \textit{B}, we created a combined feature space that concatenated all the features across both models, fit the weights and computed cross-validated $R^2$ for this combined model. Based on this combined fit and the fits of each model alone, we computed:
$R^2_{\text{unique}:A} = R^2_{\text{combined}} - R^2_{B\text{ only}}$,
$R^2_{\text{unique}:B} = R^2_{\text{combined}} - R^2_{A\text{ only}}$.


\begin{table}[t]
\centering
\captionsetup{font=normalsize} 

\caption{Summary Metrics for Pretraining and Linear Probes. SimCLR Top-1 is computed from the self-supervised contrastive objective and is not directly comparable to downstream classification. For downstream tasks, classifier checkpoints were selected by best validation Macro-F1. In-domain Top-1 accuracy is omitted because label imbalance across frames can make accuracy misleading; Macro-F1 is reported as the primary class-balanced metric. STL-10, ImageNet-100, and ImageNet-1K are treated as out-of-domain baselines because they were not pretrained on VEDB.}
\label{tab:linear_probe_summary}

\scriptsize
\setlength{\tabcolsep}{1.2pt}
\renewcommand{\arraystretch}{0.88}

\scalebox{1.00}{
\begin{tabular}{@{} p{1.25cm} p{1.65cm} r r r r @{}}
\toprule
\textbf{Task} & \textbf{Condition} & \textbf{Val Loss} & \textbf{Top-1 (\%)} & \textbf{Top-5 (\%)} & \textbf{Best Macro-F1 (\%)} \\
\midrule

\multirow{4}{*}{SimCLR}
  & Baseline    & 0.4331 & 87.60 & \multicolumn{1}{c}{--} & \multicolumn{1}{c}{--} \\
  & Fovea-Gaze  & 0.3749 & 90.43 & \multicolumn{1}{c}{--} & \multicolumn{1}{c}{--} \\
  & Periph-NF   & 0.4548 & 90.04 & \multicolumn{1}{c}{--} & \multicolumn{1}{c}{--} \\
  & Periph      & 0.4545 & 89.26 & \multicolumn{1}{c}{--} & \multicolumn{1}{c}{--} \\
\midrule

\multirow{6}{*}{In-Domain}
  & Baseline    & 0.9811 & \multicolumn{1}{c}{--} & \multicolumn{1}{c}{--} & 42.17 \\
  & Fovea-Gaze  & 1.2031 & \multicolumn{1}{c}{--} & \multicolumn{1}{c}{--} & 43.64 \\
  & Periph-NF   & 1.3090 & \multicolumn{1}{c}{--} & \multicolumn{1}{c}{--} & 30.93 \\
  & Periph      & 1.0623 & \multicolumn{1}{c}{--} & \multicolumn{1}{c}{--} & 36.56 \\
  & STL-10       & 1.6666 & \multicolumn{1}{c}{--} & \multicolumn{1}{c}{--} & 25.41 \\
  & ImageNet-100 & 1.2342 & \multicolumn{1}{c}{--} & \multicolumn{1}{c}{--} & 41.23 \\
  & ImageNet-1K & 0.9713 & \multicolumn{1}{c}{--} & \multicolumn{1}{c}{--} & 43.33 \\
\midrule

\multirow{6}{*}{VGGFace2}
  & Baseline    & 7.8101 & 5.21  & 11.73 & 3.26 \\
  & Fovea-Gaze  & 7.9104 & 4.58  & 10.76 & 2.70 \\
  & Periph-NF   & 8.0232 & 3.39  & 8.17  & 1.90 \\
  & Periph      & 8.1681 & 2.54  & 6.39  & 1.35 \\
  & STL-10       & 6.9973 & 9.55  & 18.96 & 7.43 \\
  & ImageNet-100 & 6.7985 & 10.77 & 21.07 & 8.71 \\
  & ImageNet-1K & 6.7964 & 10.74 & 21.08 & 8.77 \\
\midrule

\multirow{6}{*}{Places365}
  & Baseline    & 3.9690 & 25.63 & 51.90 & 23.16 \\
  & Fovea-Gaze  & 4.2347 & 21.86 & 46.21 & 19.14 \\
  & Periph-NF   & 4.2621 & 20.51 & 44.58 & 17.86 \\
  & Periph      & 4.2671 & 20.26 & 44.10 & 17.65 \\
  & STL-10       & 3.8281 & 26.57 & 53.47 & 24.82 \\
  & ImageNet-100 & 3.9207 & 24.99 & 51.21 & 23.32 \\
  & ImageNet-1K & 3.6264 & 30.17 & 58.46 & 28.36 \\
\bottomrule
\end{tabular}}

\end{table}

\section{Results}

\subsection{Downstream task performance}
After pretraining each model, we used linear probing to evaluate performance on in-domain frame label classification, face classification and scene classification, revealing differences in performance across conditions (Figure \ref{fig:curves-combined-all} and Table~\ref{tab:linear_probe_summary}).
In the SimCLR pretraining phase, all four VEDB conditions performed well and within a similar range, but Fovea-Gaze yielded the strongest performance (Fovea-Gaze Top-1 = 90.43\%), slightly exceeding both peripheral variants (Periph-NF = 90.04\%; Periph = 89.26\%), suggesting a small foveal advantage during representation learning. 
In the in-domain supervised probe, final validation loss was lowest for the Baseline condition, which is expected given that the Baseline images have more total information retained compared to the modified versions.
However, class-balanced Macro-F1 accuracy showed that the Fovea-Gaze model outperformed both the Baseline and the two Periph models 
(Macro-F1: Fovea-Gaze = 43.64\%, Periph-NF = 30.93\%, Periph = 36.56\%, Baseline = 42.17\%). The advantage for the Fovea-Gaze model in this task may suggest that foveally-cropped frames isolate the most diagnostic features for supporting frame category classification. The advantage for the Periph condition over the Periph-NF condition suggests that fine spatial detail in the periphery, which is removed from images via the NeuroFovea transform, provides another source of information for scene classification.

Comparing against non-VEDB baselines on the in-domain task, the VEDB Baseline and Fovea-Gaze models (Macro-F1: 42.17\%, 43.64\%) performed comparably to ImageNet-1K (43.33\%) and ImageNet-100 (41.23\%), and outperformed STL-10 (25.41\%).
This result suggests that egocentric pretraining led to representations competitive with more curated datasets for classifying naturalistic scene content.

\begin{table}[!ht]
  \begin{center}
    \caption{Top classes for in-domain classification, ranked by $\max_j|\Delta_j|$, where $\Delta_j$ is the F1 difference between two conditions. Positive values indicate condition A $>$ B. Asterisks indicate comparisons whose 95\% bootstrap CI does not overlap zero (1000-replicate parametric multinomial bootstrap; see Supp. Figure \ref{tab:TableNew_f1_diff_topk_full} for full CIs). See Supp. Figure \ref{fig:supp_frame_examples} for examples of each label and Supp. Figure \ref{fig:supp_f1_delta_bootstrap_ci} for bar-plot visualization of $\Delta$F1.}
    
    \small
    \sisetup{
      table-number-alignment=center,
      table-format=1.3,
      round-mode=places,
      round-precision=3
    }
    \setlength{\tabcolsep}{5pt}
    \renewcommand{\arraystretch}{1.05}
    \resizebox{1.00\columnwidth}{!}{%
      \begin{tabular}{r l S@{\,}c S@{\,}c S@{\,}c}
      \toprule
      {\textbf{Rank}} & {\textbf{Label}} & {\textbf{Fovea $>$ Base}} & {} & {\textbf{Fovea $>$ Periph-NF}} & {} & {\textbf{Fovea $>$ Periph}} & {} \\
      \midrule
      1  & {playing video game}   & 0.433  & {*} & 0.433  & {*} & 0.433  & {*} \\
      2  & {solving rubik's cube} & -0.116 & {*} & 0.428  & {*} & 0.018  & {} \\
      3  & {socializing}          & 0.365  & {*} & 0.318  & {*} & 0.336  & {*} \\
      4  & {computer work}        & 0.079  & {*} & 0.300  & {*} & 0.281  & {*} \\
      5  & {skateboarding}        & -0.292 & {*} & -0.170 & {*} & -0.282 & {*} \\
      6  & {playing frisbee}      & -0.251 & {*} & -0.214 & {*} & -0.263 & {*} \\
      7  & {washing dishes}       & -0.263 & {*} & -0.100 & {*} & -0.075 & {*} \\
      8  & {cooking}              & -0.221 & {*} & 0.029  & {*} & -0.078 & {*} \\
      9  & {playing pool}         & -0.200 & {*} & -0.178 & {*} & -0.194 & {*} \\
      10 & {playing cards}        & 0.178  & {*} & 0.147  & {*} & 0.178  & {*} \\
      11 & {exploring}            & 0.023  & {} & 0.124  & {*} & 0.077  & {*} \\
      12 & {playing poker}        & -0.116 & {*} & 0.053  & {*} & -0.055 & {*} \\
      13 & {preparing}            & -0.038 & {} & -0.041 & {*} & -0.094 & {*} \\
      14 & {playing sudoku}       & -0.071 & {*} & -0.048 & {*} & 0.010  & {} \\
      15 & {walking}              & -0.040 & {*} & -0.028 & {*} & -0.030 & {*} \\
      16 & {playing piano}        & 0.002  & {} & 0.002  & {} & 0.002  & {} \\
      17 & {standing}             & 0.000  & {} & 0.000  & {} & 0.000  & {} \\
      \bottomrule
      \end{tabular}%
    }
    \label{tab:TableNew_f1_diff_topk_asterisk}
  \end{center}
\end{table}

\begin{figure}[ht]
  \centering
  \includegraphics[width=1.00\columnwidth]{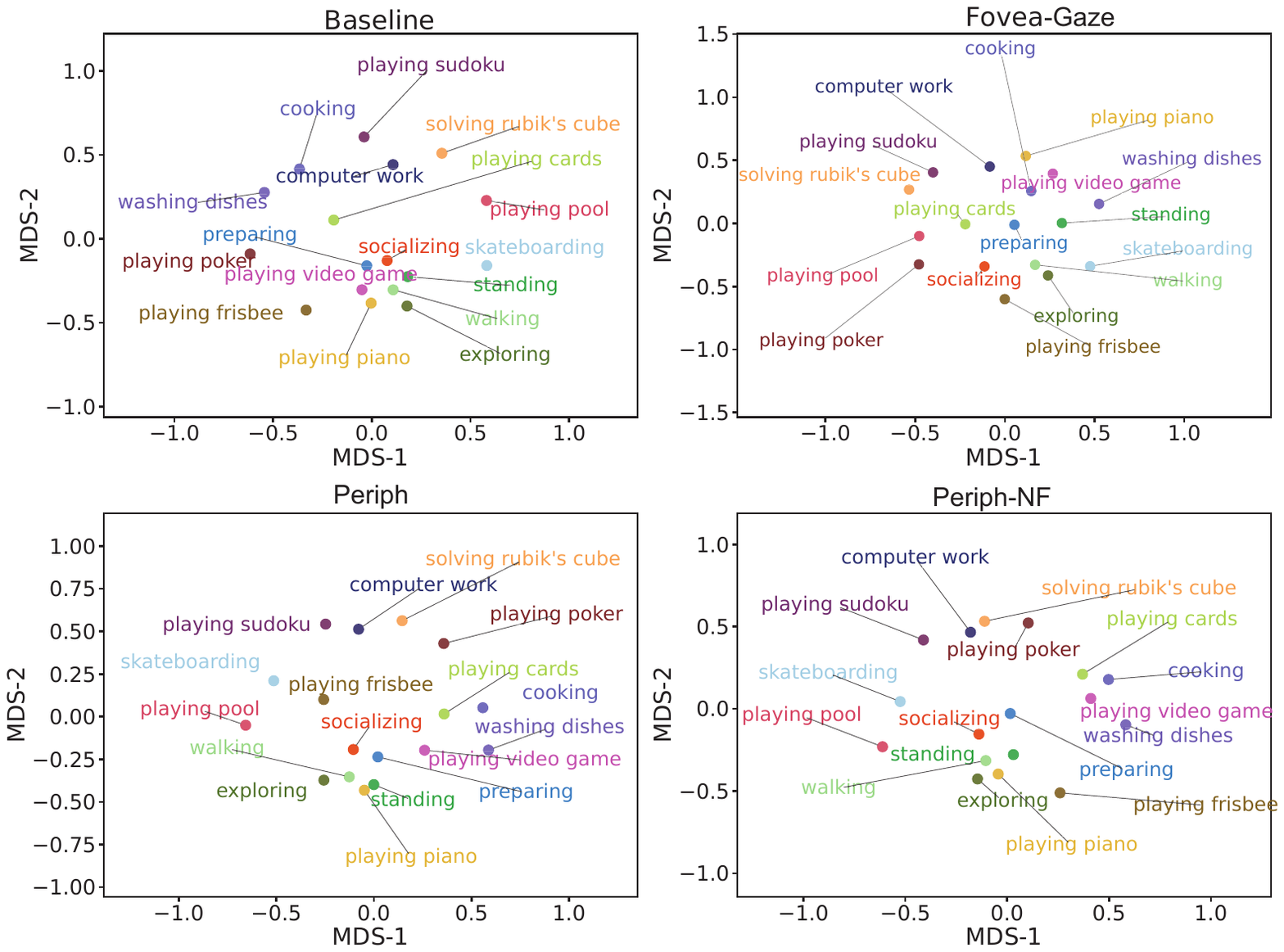}
  \caption{Multidimensional scaling (MDS) was performed using confusion patterns across the 17 task labels (see Supp. Figure \ref{fig:supp_confusion_2x2} for confusion matrices). Categories positioned closer together are more often confused by the classifier. Embeddings were aligned across conditions by a Procrustes transform.}
  \label{fig:mds_2x2}
\end{figure}

To better understand these condition differences, we next examined differences across individual categories (Table \ref{tab:TableNew_f1_diff_topk_asterisk}). When comparing Macro-F1 across conditions, we observed the Fovea-Gaze model out-performed the Periph-NF and Periph models for a subset of tasks, such as ``playing video game'', ``computer work'', and ``socializing'', while Periph-NF and Periph outperformed the Fovea-Gaze model for other tasks like ``skateboarding'', ``playing frisbee'', ``playing pool''. These results suggest a pattern in which fovea-constrained models learn more accurate representations of scene categories that include fine-scale detail or social information (screens, faces), while peripheral-constrained models learn more accurate representations of scene categories that include outdoor context or involve large-scale motor actions. 
However, there are exceptions to this suggestive pattern (e.g., ``washing dishes'' shows a peripheral advantage that does not intuitively map onto properties of that task), indicating further work is needed to build on these preliminary results.

To further explore representational differences across conditions, we performed MDS on the confusion matrices from each classifier (Figure \ref{fig:mds_2x2}; Supp. Figure \ref{fig:supp_confusion_2x2}), where categories positioned closer together are those more frequently confused. 
These plots show coarse categorical organization, such that tasks like ``cooking'', ``washing dishes'', and ``playing sudoku'' are represented similarly, while tasks like ``standing'', ``walking'', and ``exploring'' are also represented similarly.
These groupings suggest the models capture semantic or functional structure, such as stationary indoor activities versus mobile or navigational tasks. Notably, this organization emerged from self-supervised training without explicit category supervision, suggesting that the natural statistics of first-person visual experience contain sufficient structure to support categorical distinctions.

Differences in this structure are evident across conditions. For example, the ``socializing'' category in the Baseline, Periph and Periph-NF model is tightly clustered with categories like ``standing'', ``walking'' and ``preparing'', while in the Fovea-Gaze condition it is more separated from these neighboring categories. This may indicate that fovea-restricted inputs led to more effective differentiation of images containing social information (e.g., faces) from those not containing social information, which could underlie the difference in macro-F1 across conditions for the ``socializing'' label.


To provide a stronger test of these preliminary observations, we tested transfer to out-of-domain categorization of face (VGGFace2) and scene (Places365) images using a linear probe. 
Across all four VEDB conditions, we observed better overall transfer to Places365 than VGGFace2, which likely reflects the greater input distribution shift between VEDB and VGGFace2. Similarly, the non-VEDB baseline models (STL-10, ImageNet-100, ImageNet-1K) also performed better on Places365 than VGGFace2 (note also that models were not fine-tuned). The non-VEDB baselines  performed comparably to the VEDB-trained models on Places365 (Macro-F1: STL-10 = 24.82\%, ImageNet-100 = 23.32\%, ImageNet-1K = 28.36\% vs.\ VEDB Baseline = 23.16\%), but all non-VEDB baselines outperformed all VEDB-pretrained models on the VGGFace2 task, potentially reflecting limited face content in VEDB (see \textit{Discussion}).


Across the VEDB eccentricity-constrained conditions, both VGGFace2 and Places365 showed a clear fovea$>$periphery ordering: on VGGFace2, Fovea-Gaze outperformed both peripheral variants 
(Macro-F1: 2.70\% vs.\ 1.90\% and 1.35\%). 
Surprisingly, Fovea-Gaze similarly exceeded peripheral variants on Places365 
(Macro-F1: 19.14\% vs.\ 17.86\% and 17.65\%). 
This finding may be related to the semantic nature of this scene categorization task; see \textit{Discussion}.
Comparing the two peripheral conditions, Periph-NF outperformed Periph, suggesting periphery-inspired metamerization provided a small robustness benefit for out-of-domain transfer. This held for both types of transfer 
(VGGFace2 
Macro-F1: 1.90\% vs.\ 1.35\%; Places365 
Macro-F1: 17.86\% vs.\ 17.65\%). 
This is in contrast to the in-domain probe, in which Periph achieved higher Macro-F1 than Periph-NF (36.56\% vs.\ 30.93\%).

\subsection{fMRI encoding model performance}

\begin{figure}[h!t]
  \includegraphics[width=\columnwidth]{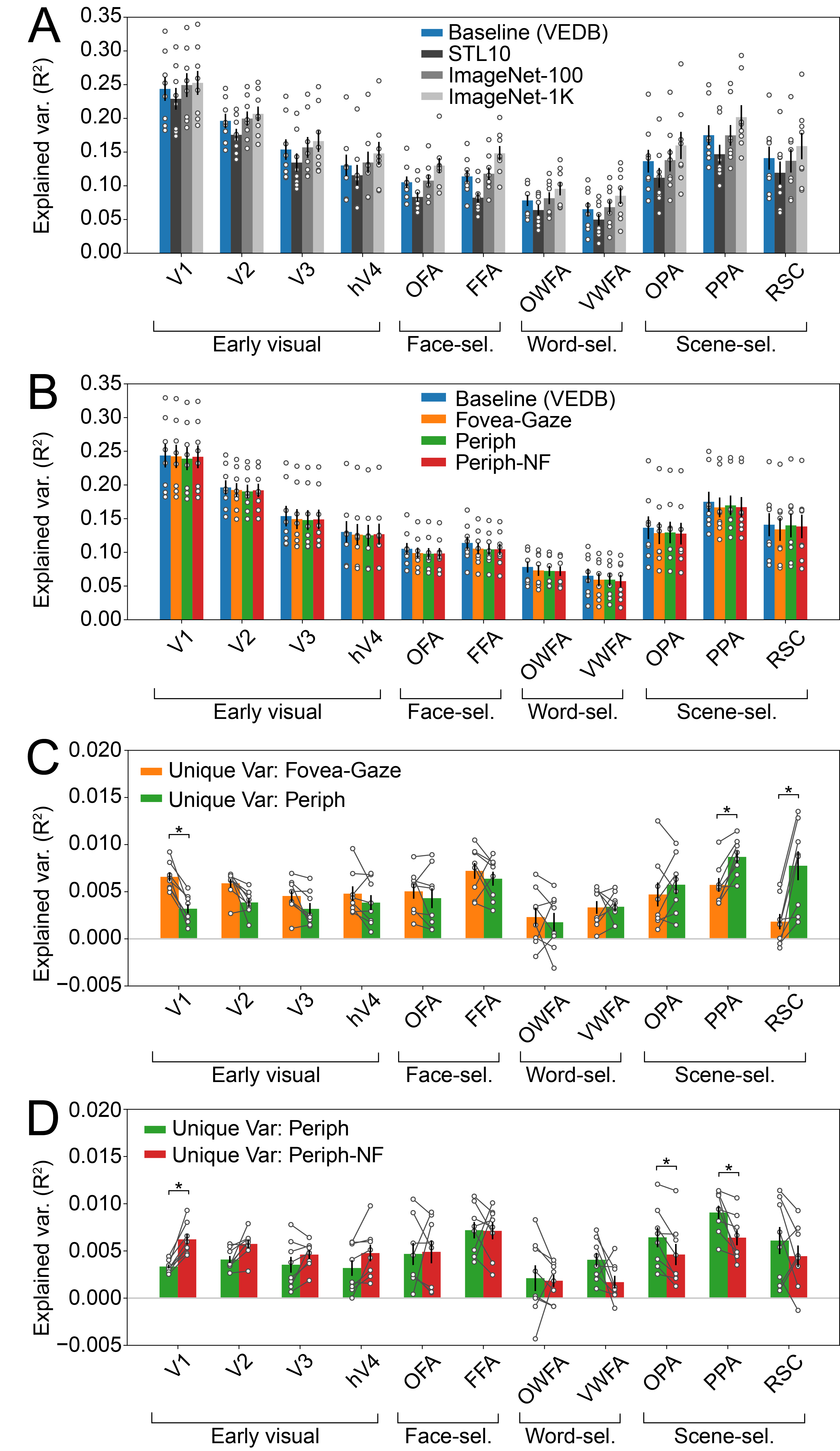}
  \caption{fMRI encoding model performance. \textbf{(A)} Voxelwise prediction accuracy (cross-validated $R^2$) for Baseline VEDB and non-VEDB comparison models, averaged across voxels in functional ROIs for each participant. \textbf{(B)} As in (A), for the four VEDB conditions. See Supp. Figure \ref{fmri-scatter-supp} for comparison of $R^2$ vs. voxelwise noise ceiling, and see Supp. Table \ref{fmri-table} for significance testing between conditions. \textbf{(C)} Unique variance for Fovea-Gaze model and Periph model. \textbf{(D)} Unique variance for Periph model and Periph-NF model. Brackets and $*$ indicate significant difference between unique variance across models (paired \textit{t}-test with permutation, all p $<$ 0.01; see \textit{Supp. Methods}). In all panels: gray dots and lines indicate individual fMRI participants, error bars show the SEM across 8 participants. Sel. = selective.}
  \label{fmri-figure}
\end{figure}

Next, we used a voxelwise encoding model framework (see \textit{Methods}) to examine alignment of each model's learned representations with the representations in target visual cortex regions.
We hypothesized that training on foveal-only inputs would lead to models more predictive of face- and word-selective cortical regions, while training on peripheral-only inputs would lead to models more predictive of scene-selective cortical regions. 
We first assessed overall model performance using the cross-validated predictive accuracy ($R^2$) of each VEDB-pretrained model and each non-VEDB comparison model (Figure \ref{fmri-figure}A-B). We observed moderately high values of $R^2$ across voxels in early and higher visual cortex regions for all of our VEDB-pretrained models, with voxelwise $R^2$ close to the noise ceiling in many voxels (Supp. Figure \ref{fmri-scatter-supp}). Across ROIs, $R^2$ for the Baseline VEDB model was comparable to that of 
the ImageNet-100 model, consistently exceeded the STL-10 model, and was lower but approaching the accuracy of the ImageNet-1K model. This difference was largest in FFA, where Baseline-VEDB $R^2$ was 0.11 $\pm$ 0.01 (mean $\pm$ SEM across 8 participants) and ImageNet-1K $R^2$ was 0.15 $\pm$ 0.01.
This suggests that pretraining with egocentric frames from VEDB, despite the more constrained and temporally correlated nature of experience-sampled data, led to comparably cortically-aligned models as pretraining on mid-sized non-egocentric datasets.

In addition, we observed differences across the different VEDB conditions, which varied across ROIs. This pattern was quantified using a two-way repeated-measures ANOVA which revealed significant main effects of ROI and model condition, as well as an interaction between ROI and model condition 
(ROI: \textit{F}(10, 70) = 43.68, p < 0.0001; 
ModelCondition: \textit{F}(3, 21) = 70.17, p < 0.0001;
ROI:ModelCondition: \textit{F}(30, 210) = 5.95, p < 0.0001;
all p-values obtained using a permutation test; see \textit{Supp. Methods}). 
Follow-up \textit{t}-tests revealed that across ROIs, all eccentricity-constrained conditions (Fovea-Gaze, Periph, and Periph-NF) explained less variance than the Baseline VEDB model (see Supp. Table \ref{fmri-table} for test statistics), consistent with the idea that cropping images reduces their total information content leading to worse neural alignment. The main exception to this was V1, in which the Fovea-Gaze and Periph-NF models performed as well as the Baseline model, though the Periph model was worse than Baseline.

Most importantly, 
consistent with our initial hypothesis, the Periph model out-performed the Fovea-Gaze model in $R^2$ for scene-selective cortical regions PPA and RSC (PPA: \textit{t}(7) = -4.86, p = 0.0074; RSC \textit{t}(7) = -4.91, p = 0.0070; paired \textit{t}-test with permutation; see Supp. Table \ref{fmri-table} for all test statistics). In face-selective and word-selective cortical regions, $R^2$ did not significantly differ between the Fovea-Gaze and the Periph model, but in primary visual cortex (V1), $R^2$ for the Fovea-Gaze model was significantly higher than for the Periph model (\textit{t}(7) = 7.04, p = 0.0062). Further confirming these differences, a variance partition analysis between the Fovea-Gaze and Periph model (Figure \ref{fmri-figure}C)  demonstrated that the Periph model explained significantly more unique variance in PPA and RSC than the Fovea-Gaze model, while the Fovea-Gaze model explained significantly more unique variance than the Periph model in V1. While numerically small, this difference between Periph and Fovea-Gaze was found for all 8 individual NSD participants for both PPA and RSC, reinforcing the reliability of this small effect. We also observed a difference between the Periph and Periph-NF conditions, such that the Periph-NF model out-performed the Periph model in V1 (\textit{t}(7) = -5.62, p = 0.0074), while the Periph model outperformed the Periph-NF model in scene-selective regions OPA and PPA (OPA: \textit{t}(7) = 3.76, p = 0.0066; PPA: \textit{t}(7) = 2.97, p = 0.0084). This pattern was confirmed with a variance partition analysis (Figure \ref{fmri-figure}D), which showed significantly more unique variance in OPA and PPA was explained by the Periph model, while more unique variance in V1 was explained by the Periph-NF model. The performance of the Periph-NF model was similar to the Fovea-Gaze model across all ROIs, with no significant differences observed between these models. 

\section{Discussion}

Natural visual inputs are highly structured, and the visual system utilizes this structure to enable adaptive behavior. In this work, we have demonstrated that CNN models trained on foveal or peripheral portions of an egocentric visual field have differential patterns of performance across categorization tasks, implying a functional difference in the statistics of these visual field portions. We also showed that self-supervised training with egocentric video frames led to comparable fMRI encoding model accuracy as training on mid-sized non-egocentric datasets, suggesting that experience-sampled egocentric data, despite its more constrained distribution, captures sufficient image statistics to support learning of cortically-aligned representations.
Finally, we demonstrated that periphery-trained models had higher predictive accuracy in scene-selective cortical regions compared to fovea-trained models. This indicates scene-selective cortical regions may selectively represent the features found in the peripheral visual field, potentially supporting behaviors such as navigation.

Across all tasks, our Fovea-Gaze model outperformed the Periph and Periph-NF models. This foveal advantage was most pronounced for face classification (VGGFace2), which is directionally consistent with our hypothesis that the central field is more informative for face processing. However, absolute performance on VGGFace2 was low across all conditions, 
potentially because VEDB pretraining has limited face content. The fovea > periphery ordering on this task should therefore be interpreted as suggestive rather than definitive. 

Surprisingly, a foveal advantage was also observed for scene classification (Places365).
One explanation for this is that the Places365 task is primarily a semantic task, which may be solvable based on diagnostic objects or materials, rather than a depth or navigational task that relies on scene geometry. This also fits with our finding that the Periph model did explain more unique variance in scene-selective cortical regions (PPA, RSC) than the Fovea-Gaze model. These scene-selective regions are thought to play key roles in encoding scene layout and other navigational information \parencite{epstein2019scene, epstein2008parahippocampal, vann2009retrosplenial}, so our results may suggest that peripheral training led to learning of more navigationally-relevant spatial information, leading to improved alignment with these areas. However, the Periph advantage in PPA and RSC was small, indicating further room for improvement in modeling scene-selective cortex.

We also observed a modest advantage for the Periph-NF model over the Periph model on out-of-distribution images, for both face and scene classification. This advantage may suggest that including the NeuroFovea transform as a precursor to self-supervised learning led to more robust model generalization performance. This is consistent with past work showing fovea-like transforms can improve robustness \citep{deza2020emergent, Wang2021}, and also with work suggesting that blurring images during model training can benefit face recogniton \citep{jang2024improved} and object categorization \citep{jinsi2023early}. In the brain, the Periph-NF model explained more unique variance in V1 than the Periph model, consistent with some past work suggesting a link between predictivity in early visual regions and model robustness \citep{li2019learning, dapello2020simulating, piper2025explicitly}, although V1 also showed an advantage in $R^2$ for the Fovea-Gaze over the Periph model. One explanation for this is that our analysis pools across all voxels regardless of retinotopic preference, so the Fovea-Gaze and Periph-NF advantages could reflect improvements in distinct voxel subpopulations.
In contrast to V1, scene-selective cortical regions (OPA, PPA) were better predicted by the Periph model than the Periph-NF model, potentially indicating that high-resolution, non-texturized peripheral images are required in order to learn a good model of scene-selective cortex. 

Our encoding model analysis did not reveal an advantage for the Fovea-Gaze model in predicting face-selective or word-selective cortical regions, as we had predicted. One potential explanation for this may be the large size of the cropping window we used in the Fovea-Gaze condition ($112\times112$ px); future work can examine the effect of a smaller cropping window 
(e.g., $10\times10$ px) to better approximate biological foveal vision. Pretraining models on foveal inputs during a supervised face or character recognition task may additionally lead to gains in prediction for face and word-selective regions.

Limitations of this current work include our use of static frames; future studies could use self-supervised video models. Future work should also investigate how foveally-trained and peripherally-trained models perform on a wider set of tasks, including more spatially-demanding scene tasks. Additionally, some downstream performance differences were modest in absolute magnitude, although we provide confidence intervals for in-domain comparisons, a full subsampling-based robustness analysis would better assess their stability; these results should be interpreted as suggestive rather than definitive. Furthermore, 
although we have shown differences between central and peripheral inputs, 
we have not yet determined which image properties underly these effects. Central and peripheral vision may differ in myriad features including spatial frequency, curvature, and semantic content; future work should aim to disentangle these factors. 

Broadly, our work has provided a step forward in understanding the link between egocentric natural image statistics, behavioral task performance, and neural representations. Understanding this link will lead to new theories on the computational and biological constraints that govern visual cortex organization.

\section{Acknowledgments and Disclosure}
D.M.D. was supported by the National Institute of General Medical Sciences through the CSUSB U-RISE Program [grant number T34GM136467], the National Science Foundation through the Cal-Bridge Program [grant number 2425075], and the NSF ExCELS Program [grant number 2322436]. This work was initiated as part of D.M.D.'s participation in the CMU/Pitt Undergraduate Program in Neural Computation at the Center for the Neural Basis of Cognition in Pittsburgh, PA, with support from the National Institute on Drug Abuse through the Interdisciplinary Training in Computational Neuroscience program [grant number 5R90DA060340-03]. D.M.D. thanks Jason F. Reimer for helpful comments and discussions throughout the study. D.M.D. and M.M.H. also thank David Pane for technical support with Carnegie Mellon University's MiND high-performance computing cluster, on which the computational experiments were performed. AI and large language models were not extensively used in the creation of this work. The authors take full responsibility for all content and declare no conflicts of interest.

\clearpage
\printbibliography

\clearpage
\section{Supp. Material}

\subsection{Supp. Methods}
\subsubsection{Frame Sampling and Processing.}
A total of 717 sessions were retrieved from the VEDB via Databrary; 203 sessions  without synchronized gaze data were excluded, yielding 514 sessions retained for processing and analysis. For each retained session, a task-relevant analysis window was defined from session logs by excluding non-task periods (e.g., setup, calibration, validation) via keyword filtering and using logged frame indices to identify task onset and offset. The window began at the final calibration segment immediately preceding task onset when available, and otherwise at the first task frame, and ended at the final task frame. Within this window, frames were sampled uniformly every 2 seconds (step size $= 2 \times \mathrm{FPS}$ at the native 25 FPS), with a maximum of 1000 frames per session. Sampled frames were decoded, converted to RGB, and processed using a deterministic pipeline consisting of a bicubic resize to 256 px followed by a $224{\times}224$ center crop. For downstream metamer generation, we additionally retained a bicubic-upsampled $512{\times}512$ version of each frame.

\subsubsection{Gaze Retrieval and Processing.}
Per-frame gaze positions were derived from the VEDB's eye-tracking recordings. First, gaze timestamps were aligned to video-frame indices. For each session, we loaded raw binocular gaze data (normalized $x$/$y$ positions and per-sample confidence values) from the VEDB's processed gaze files, along with temporal sync metadata (calibration and validation timestamp pairs) stored in the session's marker files. A first-order polynomial was fit from eye-tracker timestamps to video timestamps using these sync points, and sessions with fewer than 4 sync points or a fit $R^2 < 0.9$ were excluded. Each gaze sample was then assigned to the nearest video frame index based on the session's native frame rate. The resulting per-sample records (frame index, eye identity, normalized gaze position, and confidence) were stored as one Parquet file per session.

Second, using the frame-aligned Parquet files from the previous step, per-frame fixation locations were computed for the foveal crop construction. For each sampled video frame, all raw gaze samples from both eyes assigned to that frame were pooled and filtered by a confidence threshold ($\geq 0.30$). The per-frame gaze position was then computed as the confidence-weighted centroid over the surviving samples (e.g., $\bar{x} = \sum_i w_i x_i \,/\, \sum_i w_i$, where $w_i$ is the sample confidence). The resulting fixation coordinates, crop radius, and a flag indicating whether the gaze-based centroid or a center fallback was used, were saved as per-session CSV files at both the $224\times224$ and $512\times512$ image sizes. These fixation metadata were subsequently used for constructing the Fovea-Gaze, Periph, and Periph-NF conditions (see \textit{Methods}).

\subsubsection{Dataset Splitting Procedure.}
All dataset splits were performed at the video session level to prevent information leakage from highly correlated frames within the same recording (e.g., shared environment and temporally adjacent frames). Each sampled frame inherits a task label. For SimCLR self-supervised pretraining \citep{Chen2020}, we used the full set of sessions with synchronized gaze (514 sessions; 433,564 frames) and created an approximately 80/10/10 session split (455/28/31 train/val/test sessions) using stratified splitting to ensure balanced representation of task labels across splits. This yielded 377,462 training frames (87.06\%), 26,026 validation frames (6.00\%), and 30,076 test frames (6.94\%). Validation and test sessions were held out from all representation learning and hyperparameter selection. For downstream in-domain evaluation, we restricted the task label space to a core subset of 17 tasks meeting inclusion criteria of at least 500 frames per task and at least 7 sessions per task, yielding 298 eligible sessions (167,437 frames). These sessions were partitioned into 239/28/31 train/val/test sessions using multilabel-stratified shuffle splitting (iterative stratification) \citep{sechidis2011stratification} to preserve marginal label frequencies while maintaining strict session disjointness. This core-17 split was nested within the original session split, such that sessions retained their original train/val/test assignment after subsetting. The resulting frame counts were 116,540 train frames (69.60\%), 23,626 validation frames (14.11\%), and 27,271 test frames (16.29\%), with no sessions spanning multiple splits. For VGGFace2, the official test split contains identities disjoint from the training identities and is therefore incompatible with our closed-set identification setup. Therefore, using the official train split we created a stratified 80/20 train/validation split within each identity using a fixed random seed, enforcing at least one image in each split for identities with $\geq 2$ samples; split indices were cached to disk and reused across reruns to ensure identical partitions. Places365 underwent the same splitting protocol with a stratified 80/20 train/val split within each scene class using a fixed random seed, enforcing at least one image per split for classes with $\geq 2$ samples.

\subsubsection{NeuroFovea Metamer Generation.}
To approximate peripheral texture-tiling style representations, we generated gaze-contingent peripheral metamers using the NeuroFovea (NF) foveated style-transfer framework \citep{Deza2019}. NeuroFovea implements a feed-forward pipeline that (i) encodes an image with a VGG-based feature extractor, (ii) applies eccentricity-dependent pooling in feature space, and (iii) decodes the pooled representation back to pixel space, producing a texture-preserving metamer-like rendering. Starting from the $512\times512$ session-frame crops, we applied NF independently to each frame using scale parameter $s=0.4$ (within the model’s standard scaling regime). Pooling was centered at the per-frame gaze-defined fixation location; when fixation metadata were unavailable, we defaulted to the image center (see \textit{Supp. Methods} "Gaze Retrieval and Processing"). NF outputs were saved as PNGs in session-wise directories and subsequently resized to $224\times224$ for peripheral-view construction and SimCLR pretraining.

\subsubsection{Statistical testing of encoding model performance.}

To evaluate the significance of differences between conditions in fMRI encoding model performance, we used 
a repeated measures ANOVA test (2-way), with \textit{p}-values obtained using a non-parametric shuffling method. This was done by shuffling the values within each participant individually, which preserves the repeated measures structure of the data, and computing \textit{F}-statistics for each effect using the shuffled data, over 10,000 iterations. To obtain a \textit{p}-value for each effect, we computed the proportion of iterations where the true \textit{F}-statistic was less than or equal to the shuffled \textit{F}-statistic. This non-parametric method is appropriate for relatively small sample sizes because it is more conservative than standard parametric \textit{p}-values, and is similar to previous work \parencite{Sprague2013, Rademaker2019, Henderson2022, henderson2025dynamic}. 

We also performed paired comparisons for differences between conditions, within each ROI,
using a paired \textit{t}-test with permutation. To shuffle the data, we randomly swapped (i.e., swapped with 50$\%$ probability) the values corresponding to the two conditions for each participant, and then computed a \textit{t}-statistic for this shuffled data. This was done over each of 10,000 permutations. We obtained a two-tailed \textit{p}-value calculated based on the minimum of the proportions of permuted \textit{t}-statistics greater or less than the true \textit{t}-statistic, across iterations, multiplied by 2.

\renewcommand{\tablename}{Supp. Table}
\renewcommand{\figurename}{Supp. Figure}
\setcounter{table}{0}
\setcounter{figure}{0}

\begin{table*}[!ht]
  \begin{center}
    \caption{Linear probe hyperparameters by task. CE = cross-entropy; LS = label smoothing; pat = early-stopping patience (epochs); RLP = ReduceLROnPlateau (monitor val loss; factor = 0.1; patience = 5). In-domain used cosine LR decay with linear warmup (5 epochs). Transfer probes used early stopping (pat=10) and RLP.}
    \label{tab:supp_hparams_linear_probe_tasks}
    \small
    \setlength{\tabcolsep}{5pt}
    \renewcommand{\arraystretch}{1.10}
    \begin{tabular}{lccc}
    \toprule
    \textbf{Hyperparam.} & \textbf{In-Domain} & \textbf{VGGFace2} & \textbf{Places365} \\
    \midrule
    Backbone (frozen) & ResNet-18 & ResNet-18 & ResNet-18 \\
    Input size        & $224\times224$ & $224\times224$ & $224\times224$ \\
    Optimizer         & AdamW & AdamW & AdamW \\
    Batch size        & 256 & 60 & 60 \\
    Learning rate     & $1\times10^{-3}$ & $1\times10^{-3}$ & $1\times10^{-3}$ \\
    Weight decay      & $1\times10^{-4}$ & $1\times10^{-2}$ & $1\times10^{-2}$ \\
    Loss              & CE & CE (+ LS) & CE (+ LS) \\
    Label smoothing   & 0.0 & 0.1 & 0.1 \\
    Probe epochs      & 20 & 20 & 20 \\
    BN during probe   & eval & train & train \\
    Early stopping    & off & pat=10 & pat=10 \\
    Checkpoint select & Macro-F1 (val) & Top-1 (val) & Top-1 (val) \\
    LR schedule       & cosine + warmup & RLP & RLP \\
    Warmup epochs     & 5 & -- & -- \\
    FP16              & True & True & True \\
    Seed              & 1337 & 1337 & 1337 \\
    \bottomrule
    \end{tabular}
  \end{center}
\end{table*}
\clearpage
\begin{table*}[!ht]
  \begin{center}
    \caption{SimCLR pretraining hyperparameters. Adam = Adam optimizer; 
    NT-Xent = normalized temperature-scaled cross-entropy loss; 
    $\tau$ = softmax temperature; $p$ = probability of applying augmentation; 
    $\sigma$ = Gaussian blur strength sampled uniformly from $U(0.1, 2.0)$; 
    blur kernel size = $0.1 \times$ image size; 
    h-flip = random horizontal flip; 
    FP16 = mixed-precision training.}
    \label{tab:supp_hparams_simclr}
    \small
    \setlength{\tabcolsep}{5pt}
    \renewcommand{\arraystretch}{1.10}
    \begin{tabular}{lc}
    \toprule
    \textbf{Hyperparam.} & \textbf{SimCLR} \\
    \midrule
    Backbone          & ResNet-18 \\
    Input size        & $224\times224$ \\
    Optimizer         & Adam \\
    Batch size        & 64 \\
    Learning rate     & $3\times10^{-4}$ \\
    Weight decay      & $1\times10^{-4}$ \\
    Loss              & NT-Xent \\
    Temperature $\tau$& 0.07 \\
    Epochs            & 120 \\
    LR schedule       & fixed \\
    Proj.\ head       & Linear(512,512) $\to$ ReLU $\to$ Linear(512,128) \\
    Augmentations     & Random crop, h-flip, color jitter ($p$=0.8), \\
                      & grayscale ($p$=0.2), Gaussian blur ($\sigma \sim U(0.1,2.0)$) \\
    FP16              & True \\
    \bottomrule
    \end{tabular}
  \end{center}
\end{table*}
\clearpage

\begin{table*}[!ht]
  \begin{center}
    \caption{Comparing encoding model $R^2$ values across each pair of conditions, for each ROI. Average $R^2$ values for each ROI (across 8 participants) were compared across conditions using a paired t-test (df = 7) with permutation, see \textit{Supp. Methods}. P-values $<$ 0.01 are shown in bold.
    }
    \label{fmri-table}

\vspace{0.4cm}

\begin{tabular}{l*{8}{p{1.2cm}}}
\toprule
 & \multicolumn{2}{p{2.4cm}}{\centering Baseline vs. Fovea-Gaze} & \multicolumn{2}{p{2.4cm}}{\centering Baseline vs. Periph} & \multicolumn{2}{p{2.4cm}}{\centering Baseline vs. Periph-NF} & \multicolumn{2}{p{2.4cm}}{\centering Fovea-Gaze vs. Periph} \\
 & t & p & t & p & t & p & t & p \\
\midrule
V1 & 1.63 & 0.1452 & 5.72 & \textbf{0.0088} & 2.16 & 0.0538 & 7.04 & \textbf{0.0062} \\
V2 & 5.25 & \textbf{0.0082} & 7.15 & \textbf{0.0088} & 4.38 & \textbf{0.0084} & 3.39 & 0.0124 \\
V3 & 6.89 & 0.0112 & 10.57 & \textbf{0.0080} & 4.78 & \textbf{0.0090} & 2.80 & 0.0340 \\
hV4 & 5.71 & \textbf{0.0080} & 6.97 & \textbf{0.0066} & 5.87 & \textbf{0.0086} & 1.59 & 0.1638 \\
OFA & 6.67 & \textbf{0.0090} & 6.45 & \textbf{0.0074} & 6.14 & 0.0108 & 1.31 & 0.2054 \\
FFA & 6.45 & \textbf{0.0064} & 6.90 & \textbf{0.0096} & 6.32 & \textbf{0.0086} & 1.05 & 0.3040 \\
OWFA & 11.18 & \textbf{0.0050} & 4.94 & \textbf{0.0096} & 4.86 & \textbf{0.0078} & 0.75 & 0.4480 \\
VWFA & 5.45 & \textbf{0.0098} & 4.04 & 0.0158 & 4.72 & \textbf{0.0082} & -0.12 & 0.9344 \\
OPA & 7.86 & \textbf{0.0088} & 6.03 & \textbf{0.0086} & 8.25 & \textbf{0.0090} & -0.99 & 0.3130 \\
PPA & 9.40 & \textbf{0.0084} & 5.18 & \textbf{0.0090} & 8.39 & \textbf{0.0060} & -4.86 & \textbf{0.0074} \\
RSC & 6.80 & \textbf{0.0090} & 0.72 & 0.5054 & 2.39 & 0.0550 & -4.91 & \textbf{0.0070} \\
\bottomrule
\end{tabular}

\vspace{0.4cm}

\begin{tabular}{l*{8}{p{1.2cm}}}
\toprule
 & \multicolumn{2}{p{2.4cm}}{\centering Fovea-Gaze vs. Periph-NF} & \multicolumn{2}{p{2.4cm}}{\centering Periph vs. Periph-NF} \\
 & t & p & t & p \\
\midrule
V1 & 0.78 & 0.4580 & -5.62 & \textbf{0.0074} \\
V2 & 0.61 & 0.5154 & -3.04 & 0.0232 \\
V3 & 0.53 & 0.6186 & -1.73 & 0.1284 \\
hV4 & -0.95 & 0.3968 & -2.68 & 0.0260 \\
OFA & 0.71 & 0.4862 & -0.25 & 0.8086 \\
FFA & 0.98 & 0.3424 & -0.07 & 0.9492 \\
OWFA & 0.87 & 0.4350 & 0.15 & 0.8876 \\
VWFA & 2.31 & 0.0344 & 1.98 & 0.0970 \\
OPA & 0.77 & 0.4432 & 3.76 & \textbf{0.0066} \\
PPA & -0.46 & 0.6086 & 2.97 & \textbf{0.0084} \\
RSC & -3.37 & 0.0210 & 1.12 & 0.2682 \\
\bottomrule
\end{tabular}

    \end{center}
\end{table*}

\begin{table*}[!ht]
  \begin{center}
    \caption{Full confidence intervals for in-domain classification $\Delta$F1 (corresponding to Table \ref{tab:TableNew_f1_diff_topk_asterisk} in the main text). Classes are ranked by $\max_j|\Delta_j|$, where $\Delta_j$ is the F1 difference between two conditions. Positive values indicate condition A $>$ B. 95\% confidence intervals were estimated using a 1000-replicate parametric (multinomial) bootstrap over the aggregate confusion matrices, with row-wise resampling conditional on observed class counts. Asterisks indicate comparisons whose 95\% CI does not overlap zero. See Supp. Figure \ref{fig:supp_frame_examples} for examples of each label and Supp. Figure \ref{fig:supp_f1_delta_bootstrap_ci} for bar-plot visualization of $\Delta$F1.}
    \label{tab:TableNew_f1_diff_topk_full}
    
    \small
    \sisetup{
      table-number-alignment=center,
      table-format=1.3,
      round-mode=places,
      round-precision=3
    }
    \setlength{\tabcolsep}{4pt}
    \renewcommand{\arraystretch}{1.05}
    \resizebox{0.95\textwidth}{!}{%
      \begin{tabular}{r l S c l S c l S c l}
      \toprule
      {\textbf{Rank}} & {\textbf{Label}} & {\textbf{Fovea $>$ Base}} & {\textbf{95\% CI}} & {} & {\textbf{Fovea $>$ Periph-NF}} & {\textbf{95\% CI}} & {} & {\textbf{Fovea $>$ Periph}} & {\textbf{95\% CI}} & {} \\
      \midrule
      1  & {playing video game}   & 0.433  & {[0.399, 0.466]}   & {*} & 0.433  & {[0.401, 0.464]}   & {*} & 0.433  & {[0.399, 0.467]}   & {*} \\
      2  & {solving rubik's cube} & -0.116 & {[-0.185, -0.051]} & {*} & 0.428  & {[0.362, 0.491]}   & {*} & 0.018  & {[-0.049, 0.088]}  & {} \\
      3  & {socializing}          & 0.365  & {[0.328, 0.402]}   & {*} & 0.318  & {[0.281, 0.358]}   & {*} & 0.336  & {[0.301, 0.370]}   & {*} \\
      4  & {computer work}        & 0.079  & {[0.046, 0.116]}   & {*} & 0.300  & {[0.274, 0.326]}   & {*} & 0.281  & {[0.253, 0.309]}   & {*} \\
      5  & {skateboarding}        & -0.292 & {[-0.323, -0.261]} & {*} & -0.170 & {[-0.203, -0.139]} & {*} & -0.282 & {[-0.314, -0.253]} & {*} \\
      6  & {playing frisbee}      & -0.251 & {[-0.286, -0.215]} & {*} & -0.214 & {[-0.248, -0.181]} & {*} & -0.263 & {[-0.295, -0.229]} & {*} \\
      7  & {washing dishes}       & -0.263 & {[-0.299, -0.224]} & {*} & -0.100 & {[-0.144, -0.062]} & {*} & -0.075 & {[-0.114, -0.035]} & {*} \\
      8  & {cooking}              & -0.221 & {[-0.244, -0.200]} & {*} & 0.029  & {[0.008, 0.051]}   & {*} & -0.078 & {[-0.098, -0.054]} & {*} \\
      9  & {playing pool}         & -0.200 & {[-0.226, -0.178]} & {*} & -0.178 & {[-0.201, -0.154]} & {*} & -0.194 & {[-0.218, -0.170]} & {*} \\
      10 & {playing cards}        & 0.178  & {[0.151, 0.206]}   & {*} & 0.147  & {[0.117, 0.178]}   & {*} & 0.178  & {[0.151, 0.207]}   & {*} \\
      11 & {exploring}            & 0.023  & {[-0.040, 0.086]}  & {}  & 0.124  & {[0.072, 0.176]}   & {*} & 0.077  & {[0.023, 0.129]}   & {*} \\
      12 & {playing poker}        & -0.116 & {[-0.138, -0.094]} & {*} & 0.053  & {[0.027, 0.083]}   & {*} & -0.055 & {[-0.078, -0.032]} & {*} \\
      13 & {preparing}            & -0.038 & {[-0.074, 0.001]}  & {}  & -0.041 & {[-0.075, -0.002]} & {*} & -0.094 & {[-0.140, -0.048]} & {*} \\
      14 & {playing sudoku}       & -0.071 & {[-0.094, -0.049]} & {*} & -0.048 & {[-0.072, -0.025]} & {*} & 0.010  & {[-0.009, 0.029]}  & {} \\
      15 & {walking}              & -0.040 & {[-0.046, -0.035]} & {*} & -0.028 & {[-0.034, -0.022]} & {*} & -0.030 & {[-0.036, -0.024]} & {*} \\
      16 & {playing piano}        & 0.002  & {[0.000, 0.006]}   & {}  & 0.002  & {[0.000, 0.006]}   & {}  & 0.002  & {[0.000, 0.006]}   & {} \\
      17 & {standing}             & 0.000  & {[0.000, 0.000]}   & {}  & 0.000  & {[0.000, 0.000]}   & {}  & 0.000  & {[0.000, 0.000]}   & {} \\
      \bottomrule
      \end{tabular}%
    }
  \end{center}
\end{table*}

\begin{figure*}[!ht]
  \centering
  \vspace*{\fill}

  \includegraphics[
    width=\textwidth,
    height=0.78\textheight,
    keepaspectratio
  ]{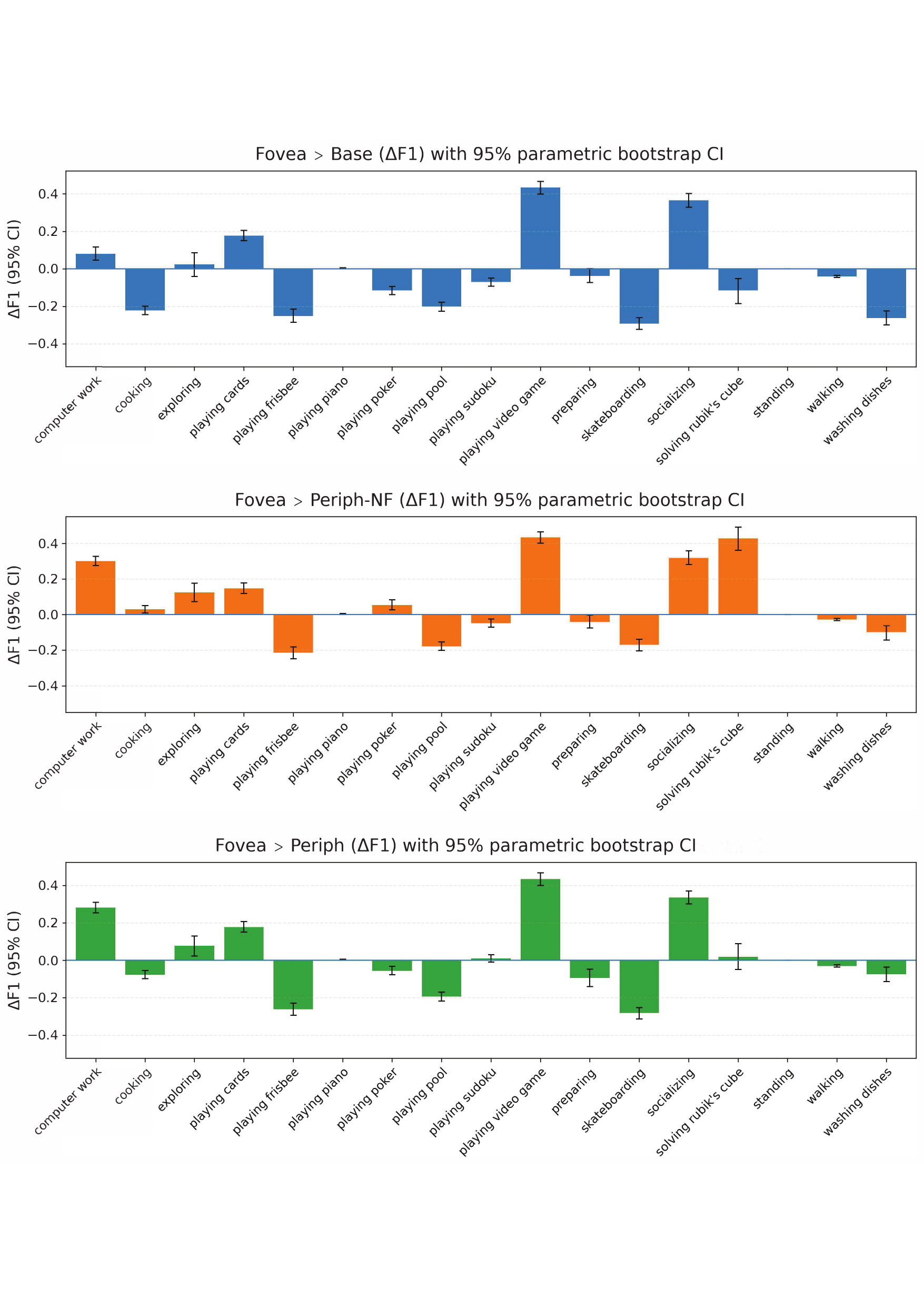}

  \captionsetup{
    width=\textwidth,
    justification=raggedright,
    singlelinecheck=false
  }
  \caption{Class-wise F1 differences ($\Delta$F1) between the Fovea condition and each comparison condition, where each panel reports $\Delta$F1 $=$ $\mathrm{F1}_{\text{Fovea}} - \mathrm{F1}_{\text{comparison}}$ (top: Fovea--Base; middle: Fovea--Periph-NF; bottom: Fovea--Periph). Error bars indicate 95\% confidence intervals from a 1000-replicate parametric (multinomial) bootstrap computed from the aggregate confusion matrices (row-wise resampling conditional on observed class counts). Positive values indicate higher F1 for Fovea; negative values indicate higher F1 for the comparison condition.}
  \label{fig:supp_f1_delta_bootstrap_ci}

  \vspace*{\fill}
\end{figure*}

\begin{figure*}[!ht]
  \centering
  \vspace*{\fill}

  \hspace*{2.7cm}%
  \begin{minipage}{0.86\textwidth}
    \centering

    \includegraphics[
      width=1.0\textwidth,
      height=0.94\textheight,
      keepaspectratio
    ]{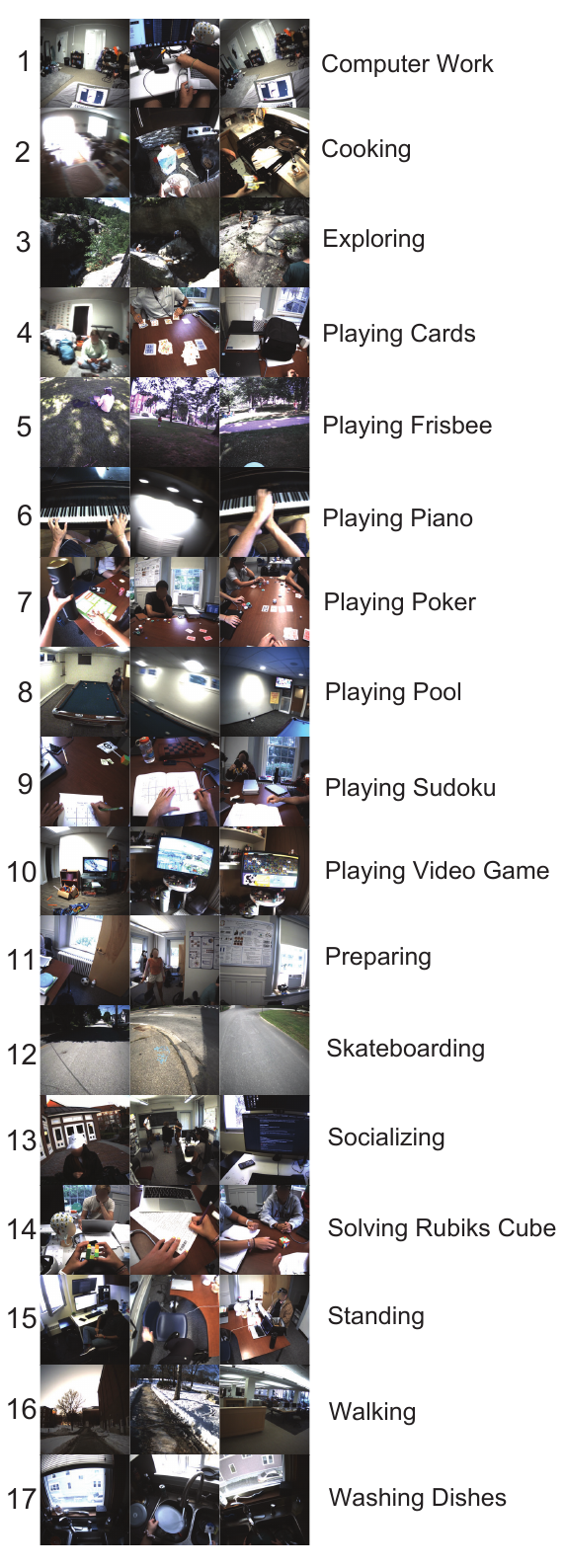}

    \captionsetup{
      justification=centering,
      singlelinecheck=true,
      margin={-2.0cm,0cm} 
    }
    \caption{Frame examples for the 17 task categories.}
    \label{fig:supp_frame_examples}

  \end{minipage}

  \vspace*{\fill}
\end{figure*}
\begin{figure*}[!ht]
  \centering
  \includegraphics[width=\textwidth]{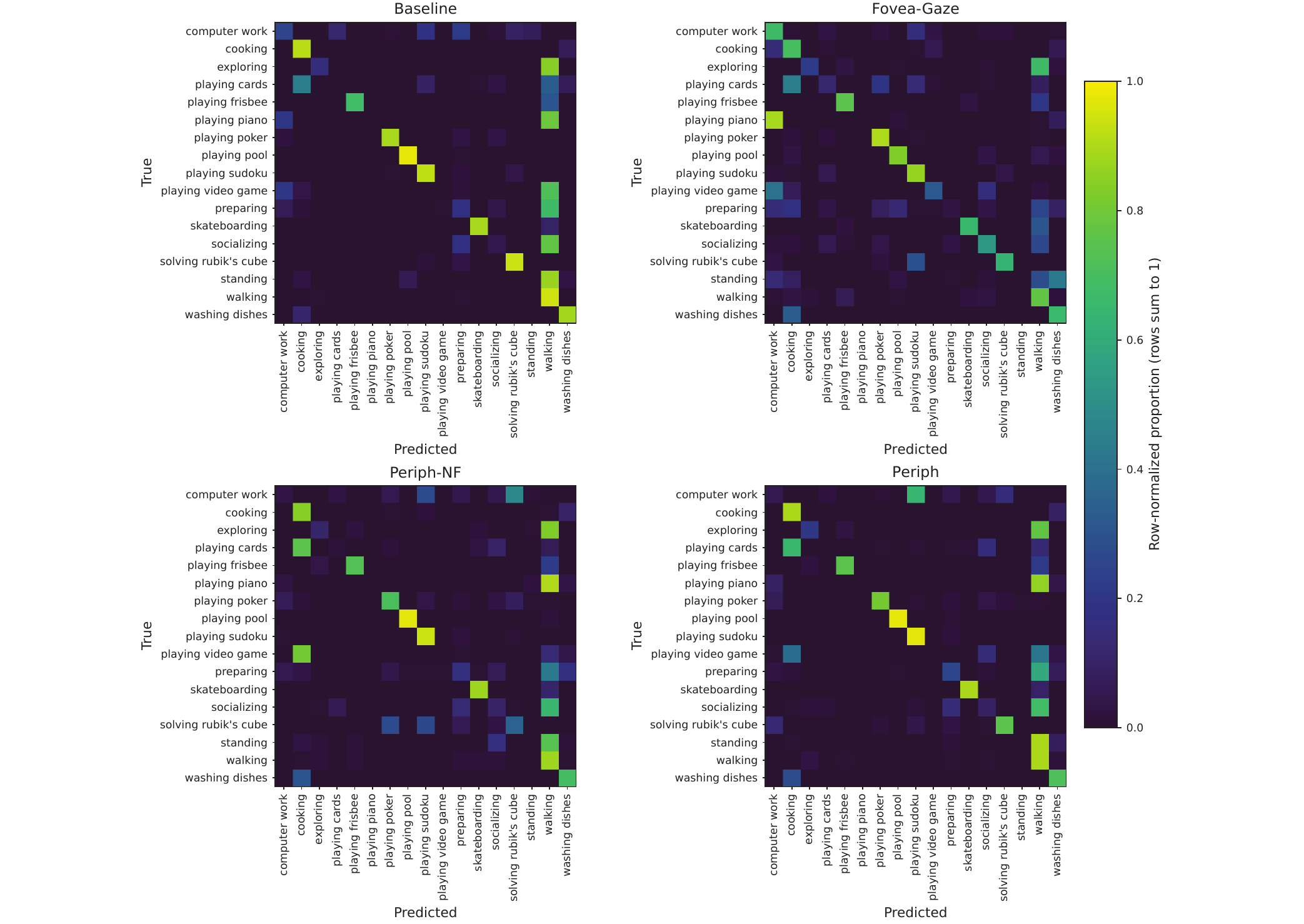}
  \caption{Row-normalized confusion matrices (rows sum to 1) for the in-domain 17-class task across conditions (Baseline, Fovea-Gaze, Periph-NF, Periph). Axes show true labels (y) and predicted labels (x).}
  \label{fig:supp_confusion_2x2}
\end{figure*}

\begin{centering}
\begin{figure*}[p]

  \includegraphics[width=\textwidth]{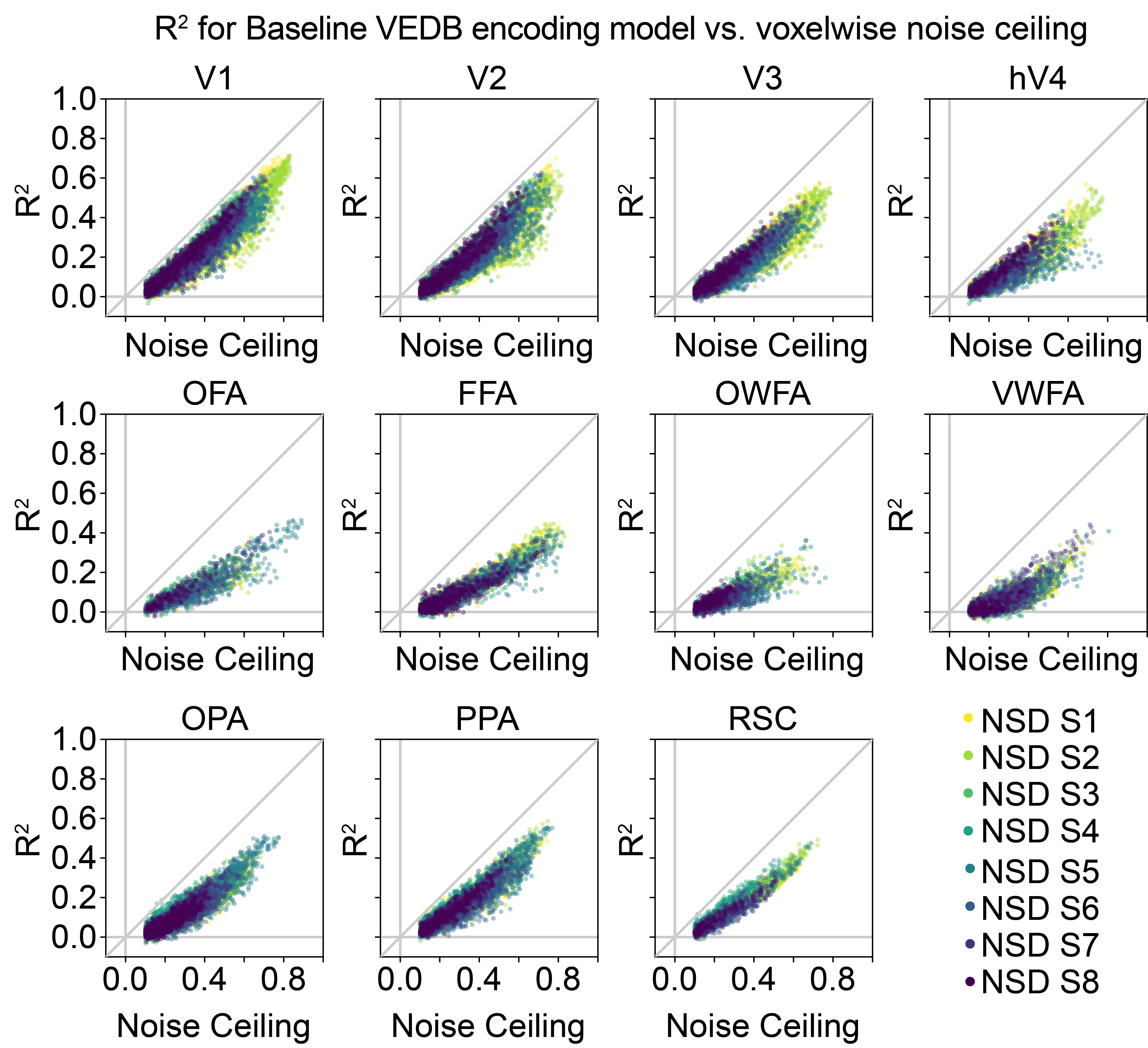}
  \caption{fMRI encoding model performance for the Baseline VEDB model condition, plotted as a function of voxelwise noise ceiling. Each panel plots voxels from a different functionally-defined ROI, each point is a voxel, colors indicate voxels from different participants. Voxelwise noise ceiling is computed with respect to beta weights averaged over 3 repetitions of each image in the NSD dataset, see \cite{Allen2022} for details.}
  \label{fmri-scatter-supp}
\end{figure*}
\end{centering}

\end{document}